\documentclass[reprint,nofootinbib,superscriptaddress,amsmath,amssymb,aps,prd]{revtex4-1}
\usepackage{amsmath,mathrsfs,amsbsy,color,graphicx,bm,amsthm,amsfonts}
\usepackage{bbm}
\usepackage{times}
\usepackage{units}
\usepackage{dcolumn}
\usepackage{graphicx}
\usepackage{epsfig}
\usepackage{epstopdf}
\usepackage[colorlinks,linkcolor=blue,anchorcolor=blue,citecolor=blue,CJKbookmarks=True]{hyperref}
\DeclareMathSymbol{\shortminus}{\mathbin}{AMSa}{"39}
\usepackage{mathrsfs}
\usepackage{braket}
\usepackage{amssymb}
\usepackage{txfonts}
\usepackage{float}
\usepackage{enumitem}
\usepackage{multirow}

\newcommand{\meq}[1]{(\ref{#1})}

\newcommand{\non}{\nonumber \\}
\newcommand{\pp}{\partial}

\begin{document}
\title{Static neutral black holes in Kalb-Ramond gravity}

\author{Wentao Liu}
\affiliation{Department of Physics, Key Laboratory of Low Dimensional Quantum Structures and Quantum Control of Ministry of Education, and Synergetic Innovation Center for Quantum Effects and Applications, Hunan Normal
University, Changsha, Hunan 410081, P. R. China}
\author{Di Wu}
\email[]{wdcwnu@163.com}
\affiliation{School of Physics and Astronomy, China West Normal University, Nanchong, Sichuan 637002, P. R. China}
\author{Jieci Wang}
\email{jcwang@hunnu.edu.cn}\affiliation{Department of Physics, Key Laboratory of Low Dimensional Quantum Structures and Quantum Control of Ministry of Education, and Synergetic Innovation Center for Quantum Effects and Applications, Hunan Normal
University, Changsha, Hunan 410081, P. R. China}

\begin{abstract}
The Kalb-Ramond (KR) gravity theory, a modified gravity theory that nonminimally couples a KR field with a nonzero vacuum expectation value for the gravitational field, can spontaneously break the Lorentz symmetry of gravity. In a recent work, Yang et al. [\href{http://dx.doi.org/10.1103/PhysRevD.108.124004}{Phys. Rev. D \textbf{108}, 124004 (2023)}] successfully derived Schwarzschild-like black hole solutions both with and without a nonzero cosmological constant within the framework of KR gravity. However, their analysis did not address the more general case of static, neutral, spherically symmetric black holes. In this paper, we fill this gap by resolving the field equations to construct more general static, neutral, spherically symmetric black hole solutions both with and without a nonzero cosmological constant. Our black hole solutions are shown to obey the first law and the Bekenstein-Smarr mass formulas of black hole thermodynamics. Moreover, we demonstrate that our static neutral spherically symmetric AdS black hole does not always satisfy the reverse isoperimetric inequality (RII), as the isoperimetric ratio can be larger or smaller than unity depending on the placement of the solution parameters within the parameter space. This behavior contrasts with the above-mentioned Schwarzschild-like AdS black hole in the KR gravity theory, which always obeys the RII. Significantly, the present more general static, neutral, spherically symmetric AdS black hole is the first example of a static AdS black hole that can violate the RII.
\end{abstract}


\maketitle

\section{Introduction}

The search for quantum gravity effects has attracted attention over the last decades. On the experimental side, we have witnessed remarkable advancements, such as gravitational wave detections produced by LIGO and Virgo \cite{LIGOScientific:2016aoc,LIGOScientific:2016sjg,LIGOScientific:2017vwq,LIGOScientific:2020iuh}, as well as high-energy particle collisions at the LHC \cite{ATLAS:2012yve,CMS:2012qbp,LHCb:2022aki}. On the theoretical side, the study of Lorentz symmetry breaking (LSB) is important for understanding quantum gravity processes in fundamental physics \cite{Kostelecky1991,Kostelecky1998,Kostelecky19891,
Gambini1999,Kostelecky2001,Tian:2021mka,Tian:2022gfa,PRD109-026004}. By investigating the low-energy contributions from Lorentz symmetry breaking, especially its impact on spacetime, and analyzing high-precision experimental data, we can explore the possibility of Lorentz symmetry breaking in spacetimes compared to general relativity (GR). This analysis also allows us to assess whether these implications are consistent with a quantum gravity model, providing an accurate description of the observed phenomena within that framework.

A notable example of a Lorentz violation model is the Bumblebee gravity theory, which includes a nonminimally coupled vector field $B_a$, known as the bumblebee field \cite{Kostelecky1989}. When this field acquires a nonzero vacuum expectation value (VEV) under an appropriate potential, it induces spontaneous LSB. Delineating a privileged direction within spacetime, the Bumblebee field implies the generation of an anisotropic energy-momentum tensor. This process extends the standard general relativity formalism by introducing new interactions, which modifies gravitational theory. The first exact bumblebee solution was introduced by Casana et al. \cite{Casana2018}. Subsequently, various spherically symmetric solutions in bumblebee gravity were found \cite{Ovgun2019,Gullu2020,
Maluf2021,Xu:2022frb,Ding2022}, while the first exact rotating bumblebee solution was derived by Poulis et al. \cite{Poulis:2021nqh}. Building on these solutions, LSB has become an active area in black hole physics \cite{Liu:2022dcn,Mai:2023ggs,Xu:2023xqh,Zhang:2023wwk,Lin:2023foj,Chen:2023cjd,Chen2020,
Wang:2021gtd,Liu:2024Lv,Mai:2024lgk,Liang:2022gdk}.

Another significant Lorentz violation model is the Kalb-Ramond (KR) gravity theory, which incorporates a nonminimally coupled tensor field $B_{ab}$, known as the KR field. This field is a tensorial field arising from the bosonic spectrum of string theory \cite{Kalb:1974yc,Kao:1996ea}. It is an antisymmetric 2-tensor, denoted as $ B_{[ab]} $, and can be conveniently decomposed as $ B_{ab}=\tilde{E}_{[a}v_{b]}+\epsilon_{abcd}v^c\tilde{B}^d $, where $ v^a $ is a timelike 4-vector. The pseudofields $ \tilde{E}_a $ and $ \tilde{B}_a $ are spacelike, satisfying $ \tilde{E}_av^a=\tilde{B}_av^a=0 $. Analogous to Maxwell electrodynamics, these fields can be interpreted as the pseudoelectric and pseudomagnetic fields, respectively. Therefore, the KR VEV yields two background vectors, in contrast to the single vector produced by the bumblebee VEV. In this framework, considerable research has been conducted on finding exact solutions in the Einstein-Kalb-Ramond field equations \cite{Chakraborty:2014fva,Maluf:2021ywn,Yang:2023wtu,Duan:2023gng}. Notably, the pioneering work by Yang et al. \cite{Yang:2023wtu} first correctly derived the Schwarzschild-like solution in this theory, and the Schwarzschild-(A)dS-like solution was also obtained by relaxing the vacuum conditions. Subsequently, the study of spacetime properties surrounding these black hole solutions has garnered significant attention in the past year, including research on black hole quasi-normal mode (QNM) frequencies \cite{Guo:2023nkd,Filho:2023ycx}, photon spheres and shadows \cite{Jha:2024xtr,
Junior:2024ety}, gravitational lensing \cite{Junior:2024vdk,2406.11582}, geodesic motion \cite{Filho:2024kbq}, etc \cite{Du:2024uhd,PRD105-103505}. However, these solutions were limited to the specific spherically symmetric scenario $(-g_{tt}=g_{rr}^{-1})$ and did not extend to the more general spherically symmetric black hole scenario $(-g_{tt}\ne g_{rr}^{-1})$. To achieve this goal, it is essential to obtain a comprehensive understanding of solutions for static neutral spherically symmetric black holes solutions in KR gravity theory.

In this paper, we revisit the field equations in KR gravity and propose new static neutral black hole solutions, which differ from the solution in Ref. \cite{Yang:2023wtu} by exhibiting a more general spherically symmetric form. Afterwards, we calculate the relevant thermodynamic quantities and find that the new solutions satisfy the first law of thermodynamics and the Bekenstein-Smarr relation simultaneously. We also investigate the reverse isoperimetric inequality (RII) of the solutions and find that our new AdS black hole solution exhibits violations of the RII. Remarkably, this is the first instance where a static AdS black hole can violate RII. In contrast, the Schwarzschild-like AdS black hole solution given in Ref. \cite{Yang:2023wtu} always obeys the RII.

The remaining part of this paper is organized as follows. In Sec. \ref{Sec.2}, we provide a concise overview of the Kalb-Ramond gravity theory. We solve the equations of the theory to obtain black hole exact solutions for two types of static neutral spacetimes (the cases where $-g_{tt}=g_{rr}^{-1}$ and $-g_{tt}\ne g_{rr}^{-1}$), considering both with and without the cosmological constant. In Sec. \ref{Sec.3}, some basic thermodynamic properties of the obtained Kalb-Ramond black holes are analyzed. In Sec. \ref{Sec.4}, we explore the dynamical evolution of the scalar field for both types of spacetimes. Both types of solutions decay rapidly with an increase in the Lorentz-violating parameter, with the latter type being more sensitive. Finally, our conclusions and outlooks are given in Sec. \ref{Sec.5}.

\section{KR black hole solutions}\label{Sec.2}

Inspired by the gravitational sector of the Standard-Model Extension (SME), one considers a self-interacting potential for the KR field, which has a non-vanishing VEV, i.e., $ \langle B_{ab}\rangle =\beta_{ab} $ \cite{Altschul:2009ae}. To this end, it is assumed a potential $ V $ of the form
\begin{equation*}
V=V(B_{ab}B^{ab}\pm b^2).
\end{equation*}
The sign $ \pm $ is chosen to ensure that $ b^2 $ is a positive constant. Consequently, the VEV is determined by the constant norm condition $ \beta^{ab}\beta_{ab}=\mp b^2 $. This assumption leads to the spontaneous breaking of Lorentz symmetry due to the self-interaction of the KR field \cite{Lessa:2019bgi}. Furthermore, the gauge invariance $ B_{ab}\rightarrow B_{ab}+\partial_{[a}\Gamma_{b]} $ of the KR field is spontaneously broken \cite{Yang:2023wtu}.

Let us consider the Einstein-Hilbert action nonminimally coupled to a self-interacting KR field \cite{Altschul:2009ae}, as
\begin{equation}
\begin{aligned}\label{action}
\mathcal{S}=&\frac{1}{2\hat{\kappa}}\int d^4x\sqrt{-g}\bigg[R-2\Lambda-\frac{1}{6}H^{abc}H_{abc}-V(B^{ab}B_{ab})\\
&+\xi_2B^{ca}B^{b}{}_{a}R_{cb}+\xi_3B^{ab}B_{ab}R\bigg]+\int d^4x\sqrt{-g}\mathcal{L}_\text{M},
\end{aligned}
\end{equation}
where $ \hat{\kappa} = 8\pi G_N $ is the gravitational coupling constant. Here, $ \Lambda $ is the cosmological constant and $ \xi_2 $ and $ \xi_3 $ are the real coupling constants which control the nonminimal gravity interaction with the KR field. $ H_{abc} $ is the totally antisymmetric field-strength tensor, defined by
\begin{align*}
H_{abc}\equiv\partial_aB_{bc}+\partial_b B_{ca}+\partial_c B_{ab}.
\end{align*}
The potential $ V $, chosen to ensure a nonzero VEV for the KR field and to be zero at its minimum, triggers spontaneous Lorentz symmetry breaking. It is worth noting that the term $ \xi_3 B^{ab}B_{ab}R $ in the action \meq{action} transforms to $ \mp \xi_3 b^2 R $ in the vacuum. This transformation allows the term to be absorbed into the Einstein-Hilbert action through a redefinition of variables.

By varying the action \eqref{action} with respect to the metric $ g_{ab} $, we obtain the following gravitational field equations:
\begin{align}
R_{ab}-\frac{1}{2}g_{ab}R+\Lambda g_{ab}=T^\text{KR}_{ab}+T^\text{M}_{ab},
\end{align}
where $ T^\text{M}_{ab} $ is the energy-momentum tensor of matter fields, and
\begin{equation}
\begin{aligned}
T^\text{KR}_{ab}=&\frac{1}{2}H_{acd}H_{b}{}^{cd} -\frac{1}{12}g_{ab}H^{cde}H_{cde}+2V'(X)B_{ca}B^{c}{}_{b} \\
&-g_{ab}V(X) +\xi_2\bigg[\frac{1}{2}g_{ab}B^{ce}B^{d}{}_{e}R_{cd} -B^{c}{}_{a}B^{d}{}_{b}R_{cd} \\
&-B^{cd}B_{bd}R_{ac} -B^{cd}B_{ad}R_{bc} +\frac{1}{2}\nabla_{c}\nabla_{a}(B^{cd}B_{bd}) \\
&+\frac{1}{2}\nabla_{c}\nabla_{b}(B^{cd}B_{ad}) -\frac{1}{2}\nabla^c\nabla_c(B_{a}{}^{d}B_{bd}) \\
&-\frac{1}{2}g_{ab}\nabla_{c}\nabla_{d}(B^{ce}B^{d}{}_{e})\bigg],
\end{aligned}
\end{equation}
which can be considered as the energy-momentum tensor of the KR field. The prime indicates the derivative with respect to the variable of the corresponding functions.
On the other hand, by varying the action \meq{action} with respect to the KR field $ B_{ab} $, we can obtain the equation of motion for the KR field, which is given by $ \Pi_{ab}=0 $, where
\begin{align}
\Pi_{ab} = \nabla^c H_{abc} + 3 \xi_2 R_{ca} B^c_{~b} - 3 \xi_2 R_{cb} B^c_{~a} - 6 V' B_{ab}.
\end{align}

Under the VEV configuration, i.e., $ B_{ab}B^{ab}=\beta_{ab}\beta^{ab} $, we can consider the differential form \cite{Lessa:2019bgi}
\begin{align*}
\beta_2=-\tilde{E}(r)dt \wedge dr.
\end{align*}
The only nonvanishing terms are $ \beta_{rt}=-\beta_{tr}=\tilde{E}(r) $ in the VEV. Equivalently, in terms of matrix forms:
\begin{equation*}
\beta_{ab}=\left(
\begin{array}{cccc}
0 & -\tilde{E}(r) & 0 & 0 \\
\tilde{E}(r) & 0 & 0 & 0 \\
0 & 0 & 0 & 0 \\
0 & 0 & 0 & 0
\end{array}\right),
\end{equation*}
it follows that the vacuum field exhibits a pseudoelectric configuration.
Therefore, this configuration automatically causes the KR field strength to vanish, i.e., $ H_{abc}=0 $ or $ H_3=d\beta_2=0 $ \cite{Yang:2023wtu}.
Regarding the potential $ V $, a prime example that meets these conditions is a smooth quadratic function, similar to the one proposed by Casana et al. \cite{Casana2018}, given by:
\begin{align}\label{VV}
V=V(X)=\frac{1}{2}\lambda X^2,
\end{align}
where $\lambda$ is a constant and $X$ represents a generic potential argument.
Consequently, the VEV, $ \beta_{ab} $, arises as a solution of $ V=V'=0$.
However, this assumption is only applicable in the special case where the cosmological constant is not considered.
To obtain asymptotically (A)dS spacetime in this theory, that is, when a nonzero cosmological constant exists, consider relaxing the vacuum conditions.
Similar to the potential given by Maluf et al. \cite{Maluf2021}, another simple choice is a linear function defined as:
\begin{align}
V=V(\lambda,X)=\lambda X.
\end{align}
Here, $ \lambda $ is interpreted as a Lagrange-multiplier field \cite{Bluhm:2007bd}.
Its equation of motion ensures the vacuum condition $ X=0 $, leading to $ V=0 $ for any field $ \lambda $ on-shell.
Notably, this form of potential manifests itself as a cosmological constant.
Thus, the potential and its first-order derivative satisfy
\begin{align}\label{VVV}
&V(\beta^{ab} \beta_{ab}+b^2)=\lambda(\beta^{ab}\beta_{ab}+b^2)=0,\\ \label{Vp}
&V'(\beta^{ab}\beta_{ab}+b^2)=\lambda,
\end{align}
where $ V'(X)=dV(X)/dX $.
Then, we can define the efficient gravitational field equation, satisfying $\mathcal{G}_{ab}=0$, as follows:
\begin{align}\label{EQG}
\mathcal{G}_{ab} = R_{ab} -\Lambda g_{ab} -\xi_2\mathcal{B}_{ab} -\lambda(2\beta_{ac} \beta_{b}{}^{c}+b^2g_{ab}),
\end{align}
with
\begin{equation}\label{EQB}
\begin{aligned}
\mathcal{B}_{ab} = &g_{ab}\beta^{ce}\beta^{d}{}_{e}R_{cd} -\beta^{c}{}_{a}\beta^{d}{}_{b}R_{cd} -\beta^{cd}\beta_{ad}R_{bc} \\
&-\beta^{cd}\beta_{bd}R_{ac} +\frac{1}{2}\nabla_c\nabla_{a}(\beta^{cd}\beta_{bd}) \\
&+\frac{1}{2}\nabla_c\nabla_{b}(\beta^{cd}\beta_{ad}) -\frac{1}{2}\nabla^c\nabla_c(\beta_a{}^{d}\beta_{bd}).
\end{aligned}
\end{equation}
The details of the covariant derivative, as presented in Appendix \ref{AppendixA}, are provided.

In this work, we assume the metric corresponds to a general spherical symmetry solution and adopt the following line element:
\begin{equation}
\begin{aligned}
ds^2=&-A(r)dt^2+B(r)dr^2+r^2d\theta^2+r^2\sin^2\theta d\varphi^2.
\end{aligned}
\end{equation}
The form corresponding to the above metric for the pseudoelectric field $ \tilde{E}(r) $ is given by:
\begin{align}
\tilde{E}(r)=|b|\sqrt{\frac{A(r)B(r)}{2}},
\end{align}
so that the constant norm condition $ \beta^{ab}\beta_{ab}=-b^2 $ is satisfied. Then, the nonzero components of the efficient gravitational field equations, associated with the metric, are
\begin{align} \label{E00}
&\mathcal{G}_{tt}=\frac{(1-\ell)A}{4rB}\left(4A'/A-r\varUpsilon \right)+\Lambda~A,\\  \label{E11}
&\mathcal{G}_{rr}=\frac{(1-\ell)}{4r}\left(4B'/B+r\varUpsilon \right)-\Lambda~B,\\
&\begin{aligned} \label{E22}
\mathcal{G}_{\theta\theta}=&\frac{1}{2B}\left[\ell r^2\varUpsilon-r(1+\ell)(A'/A-B'/B)+2\ell-2  \right]\\
&+1-r^2b^2\lambda-r^2\Lambda,
\end{aligned}
\end{align}
and $ \mathcal{G}_{\varphi\varphi}=\mathcal{G}_{\theta\theta}\sin^2\theta $, where $ \varUpsilon \equiv(A'/A)^2-2A''/A+A'B'/(AB) $ and $\ell\equiv \xi_2 b^2/2$.
The nonzero components of the equations of motion for the KR field are give by
\begin{equation}
\Pi_{tr}=6\lambda\tilde{E}(r)+\frac{3\xi_2\tilde{E}(r)}{rB}\left[A'/A-B'/B-r\varUpsilon/2\right],
\end{equation}
and $ \Pi_{rt}=-\Pi_{tr} $.
Given $ \Pi_{ab}=0 $, we can thus find that the specific form of the pseudoelectric field $ \tilde{E}(r) $ does not affect the equations of motion for the KR field.
It is worth noting that if we consider the Casana condition \meq{VV} with $ \Lambda=0 $, the non-zero components of the KR field equations can be obtained from the combinations of the effective gravitational field equations, as shown below:
\begin{align}\label{Pitr}
\Pi_{tr}=\frac{3\xi_2 \tilde{E}(r)}{1-\ell}\left(\mathcal{G}_{tt}/A-\mathcal{G}_{rr}/B\right).
\end{align}
Even considering the Maluf condition \meq{VVV} with $ \Lambda\neq 0 $, the equation \meq{Pitr} still holds when $ \Lambda=(1-\ell)\lambda/\xi_2 $, which is a necessary condition for this theory to handle asymptotically (A)dS cases \cite{Yang:2023wtu}.
Therefore, in the following discussion, we can focus solely on the gravitational field equations.

\subsection{Asymptotically flat spacetime}

We first consider asymptotically flat spacetimes, which do not include a cosmological constant, and aim to construct two different types of static neutral spherically symmetric black hole solutions within KR gravity.
In this case, we need to assume both the cosmological constant $ \Lambda = 0 $ and the Lagrange-multiplier field $ \lambda = 0 $.
Then, using Eqs. \meq{E00} and \meq{E11}, we can obtain a simple relation:
\begin{align}
\mathcal{G}_{tt}+\frac{A}{B}\mathcal{G}_{rr}=\frac{1-\ell}{rB^2}\pp_r(AB)=0.
\end{align}
In other words,
\begin{align}\label{BBB}
AB=\mathcal{C}_1 \Rightarrow B(r)=\frac{\mathcal{C}_1}{A(r)},
\end{align}
where $ \mathcal{C}_1 $ is an arbitrary constant. At this point, the relation $ G_{\theta\theta}=0 $ gives the following equation involving only the metric function $A(r)$:
\begin{align}\label{Apd}
\left(\ell \pp_r^2+\frac{1+\ell}{r}\pp_r+\frac{1-\ell}{r^2}\right)A(r)-\frac{\mathcal{C}_1}{r^2}=0,
\end{align}
The solution of the above equation is:
\begin{align}\label{caseAB}
A(r)=\frac{\mathcal{C}_1}{1-\ell}+\frac{\mathcal{C}_2}{r}+\frac{\mathcal{C}_3}{r^{\ell^{-1}-1}},
\end{align}
where $ \mathcal{C}_2 $ and $ \mathcal{C}_3 $ are integration constants.
To satisfy the equations $ G_{tt}=G_{rr}=0 $ without fixing the Lorentz violation parameter $ \ell $, we must set $ \mathcal{C}_3=0 $.
The constant $ \mathcal{C}_2 $ can be interpreted as the black hole mass parameter by $ \mathcal{C}_2=-2M $.
In order to ensure that our solution reduces to the Schwarzschild solution when $ \ell=0 $, there are two different choices for the constant $ \mathcal{C}_1 $.

\textbf{Case A:}
We consider a general spherically symmetric structure. By setting $ \mathcal{C}_1=1-\ell $, we can rewrite Eq. \meq{caseAB} as follows:
\begin{align}
A(r)=1-\frac{2M}{r},
\end{align}
which is the same as the Schwarzschild black hole case. However, the metric function $ B(r)=(1-\ell)A(r)^{-1} $ ensures that the spacetime structure is different from that of the Schwarzschild black hole case. Therefore, the general static neutral spherically symmetric black hole solution can be obtained as:
\begin{equation}\label{dscaseA}
\begin{aligned}
ds^2_\text{A}=&-\left(1-\frac{2M}{r}\right)dt^2+\frac{1-\ell}{\left(1-\frac{2M}{r}\right)}dr^2\\
&+r^2d\theta^2+r^2\sin^2\theta d\varphi^2.
\end{aligned}
\end{equation}
It is straightforward to verify that the above solution \meq{dscaseA} satisfies all field equations \meq{EQG} when $ \Lambda=\lambda=0 $. The dimensionless parameter $ \ell $ represents the impact of Lorentz violation on spacetime, which is caused by the VEV of the KR field. Note that because the Riemann tensor is given by
\begin{align}
R_{\mu\nu}=\text{diag}\left\{0,0,\frac{\ell}{\ell-1},\frac{\ell}{\ell-1}\sin^2\theta \right\},
\end{align}
as $ r\rightarrow \infty $, the spacetime is not asymptotically Minkowski. Furthermore, we computed the Kretschmann scalar
\begin{align}
R^{abcd}R_{abcd}=\frac{4\left(12M^2-4Mr\ell+r^2\ell^2\right)}{r^6(1-\ell)^2},
\end{align}
which differs clearly from the Schwarzschild Kretschmann invariant when $ \ell $ is nonzero. This implies that Lorentz-violating effects cannot be eliminated by coordinate transformations. At $ r=2M $, the Kretschmann invariant remains finite, suggesting that this singularity can be eliminated through a suitable coordinate transformation. In contrast, the Kretschmann invariant at $ r=0 $ is infinite, indicating the intrinsic singularity that cannot be removed. Thus, the fundamental nature of the singularities at $ r=0 $ and $ r=2M $ (the event horizon) is preserved.

\textbf{Case B:}
We consider a special spherically symmetric structure. By setting $ \mathcal{C}_1=1 $, we can rewrite Eq. \meq{caseAB} as follows:
\begin{align}
A(r)=\frac{1}{1-\ell}-\frac{2M}{r}.
\end{align}
Using the relation in Eq. \meq{BBB}, we can obtain the special static neutral spherically symmetric black hole solution as
\begin{equation}\label{dscaseB}
\begin{aligned}
ds^2_\text{B}=&-\left(\frac{1}{1-\ell}-\frac{2M}{r}\right)dt^2+\frac{1}{\frac{1}{1-\ell}-\frac{2M}{r}}dr^2\\
&+r^2d\theta^2+r^2\sin^2\theta d\varphi^2.
\end{aligned}
\end{equation}
This metric was already given in Ref. \cite{Yang:2023wtu} and also characterizes the effect of the Lorentz-violating parameter $ \ell $ on spacetime, but its properties are distinctly different from those of Case A. They have different black hole event horizon and different Kretschmann scalar. The event horizon for the metric of Case B is given by $ r_h=2(1-\ell)M $, which results in a smaller horizon radius compared to Case A if the Lorentz-violating parameter $ \ell > 0 $.

\subsection{Asymptotically (A)dS spacetime}
If we are to consider the effects of the cosmological constant, we must also consider a non-vanishing Lagrange-multiplier field.
In this cases, it is evident that the relation given by Eq. \meq{BBB} still holds. 
By setting the metric function $ B(r)=\mathcal{C}_1 A(r)^{-1} $, the equation induced by the relation $ \mathcal{G}_{\theta\theta}=0 $ is rewritten as
\begin{align}\label{ApdL}
\left(\ell \pp_r^2+\frac{1+\ell}{r}\pp_r+\frac{1-\ell}{r^2}\right)A(r)-\frac{\mathcal{C}_1}{r^2}(1-r^2b^2\lambda-r^2\Lambda)=0.
\end{align}
For Case A, i.e., $ \mathcal{C}_1=1-\ell $, the metric function $A(r)$ is given by
\begin{align}
A(r)=1-\frac{2M}{r}-\frac{(1-\ell)(b^2\lambda+\Lambda)}{3(1+\ell)}r^2.
\end{align}
To ensure $ G_{tt}=G_{rr}=0 $, the cosmological constant $ \Lambda $ and the Lagrange-multiplier field $ \lambda $ need satisfy the following relation:
\begin{align}\label{eLl}
2\ell \Lambda-(1-\ell)b^2\lambda=0.
\end{align}
We can define an effective cosmological constant $ \Lambda_e = \lambda/\xi_2 $, thus Eq. \meq{eLl} can be rewritten as:
\begin{align}\label{eLl1}
\Lambda = (1-\ell)\Lambda_e.
\end{align}
The constraint \meq{eLl1} is a necessary condition to generate the exact solution with a cosmological constant from the efficient gravitational field equation \meq{EQG} and the potential \meq{Vp}.

Then, the corresponding (A)dS extension for Case A is
\begin{equation}
\begin{aligned}\label{KRAdS}
ds^2=&-\left[1-\frac{2M}{r}-\frac{(1-\ell)\Lambda_e}{3}r^2\right]dt^2\\
&+\frac{(1-\ell)dr^2}{1-\frac{2M}{r}-\frac{(1-\ell)\Lambda_e}{3}r^2}+r^2d\theta^2+r^2\sin^2\theta d\varphi^2.
\end{aligned}
\end{equation}
The general spherically symmetric solution \meq{KRAdS} can be verified to satisfy field equation \meq{EQG} when $ \Lambda = (1-\ell)\Lambda_e $. When $ \Lambda_e=0 $, this solution reduces to the case given in Eq. \meq{dscaseA} in the last subsection. Similarly, by setting the Lorentz-violating parameter $ \ell=0 $, the solution recovers the Schwarzschild-(A)dS black hole cases.

On the other hand, for Case B ($ \mathcal{C}_1=1 $), the solution for differential equation \meq{ApdL} is given by
\begin{align}
A(r)=\frac{1}{1-\ell}-\frac{2M}{r}-\frac{(b^2\lambda+\Lambda)}{3(1+\ell)}r^2.
\end{align}
Applying the same constraint \meq{eLl1}, we obtain the asymptotically (A)dS solutions for Case B, yielding
\begin{equation}
\begin{aligned}
ds^2=&-\left[\frac{1}{1-\ell}-\frac{2M}{r}-\frac{\Lambda_e}{3}r^2\right]dt^2\\
&+\frac{1}{\frac{1}{1-\ell}-\frac{2M}{r}-\frac{\Lambda_e}{3}r^2}dr^2+r^2d\theta^2+r^2\sin^2\theta d\varphi^2 \, ,
\end{aligned}
\end{equation}
which is consistent with the solution given in Eq. (17) in Ref. \cite{Yang:2023wtu}.

\begin{figure}[t]
\centering
\includegraphics[width=0.9\linewidth]{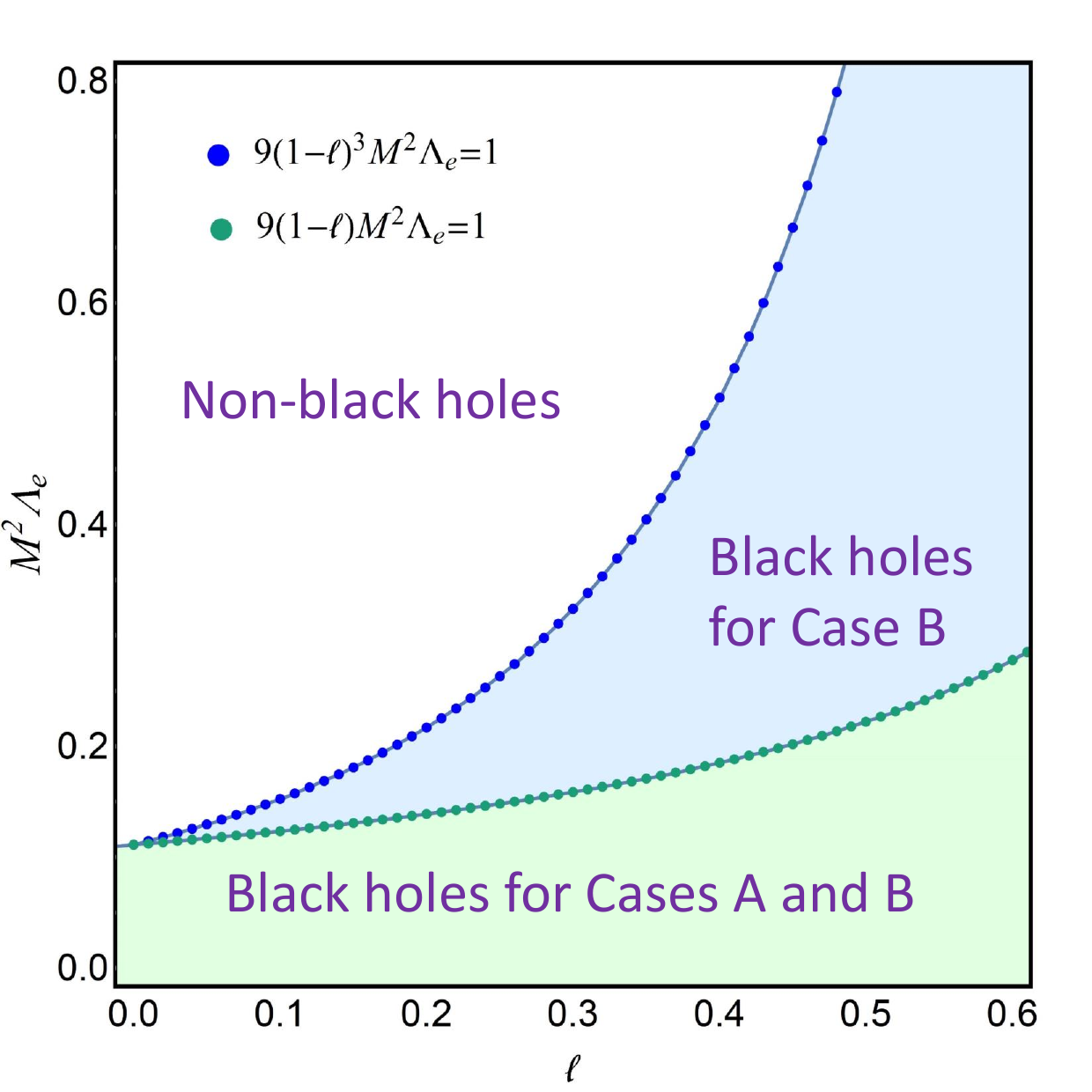}
\caption{The parameter space $ (\ell, M^2\Lambda_e) $ for black hole solutions, where the region in green represents the parameter space for Case A, while the region in blue (including green) represents that for Case B.}
\label{fig1}
\end{figure}

In asymptotically (A)dS spacetimes, the two Cases have different Ricci tensors at infinity, which contrasts with those of the asymptotically flat solutions. However, the additional effect of the cosmological constant on the Kretschmann scalar is the same in both Cases, and it has the following form:
\begin{align}
R_{ab}^{\text{(A)dS}}-R_{ab}^{\text{flat}}=\frac{8}{3}\left[\Lambda_e^2-\frac{\Lambda_e\ell}{r^2(1-\ell)}\right],
\end{align}
and the Ricci tensors remain finite at their event horizons, respectively.
The function $ A(r) $ can be expressed in terms of the dS black hole horizons as follows:
\begin{align}
A(r)=(1-\ell)^{(s-1)^2}\frac{\Lambda_e}{3}\left(1-\frac{r_h}{r}\right)\left(r_c-r\right)\left(r+r_h+r_c \right),
\end{align}
where $s = 0$ for Case A, $s = 1$ for Case B,  $ r_h $ and $ r_c $ represent the event horizon and cosmological horizon, respectively. When the effective cosmological constant is negative ($ \Lambda_e<0 $), an event horizon forms regardless of the parameter values. Therefore, both Case A and Case B metrics invariably support black hole solutions under this condition. However, as shown in Fig. \ref{fig1}, when the effective cosmological constant is positive, i.e., $ \Lambda_e > 0 $, the black hole solutions exist only for the parameters $ \ell $ and $  M^2\Lambda_e $ satisfying: $ 9(1-\ell) M^2\Lambda_e\geq1 $ for Case A and $ 9(1-\ell)^3 M^2\Lambda_e\geq1 $ for Case B, respectively,  where the equality indicates the event horizon and the cosmological horizon coincide.

\section{Thermodynamics}\label{Sec.3}

\subsection{Thermodynamical quantities}
We now proceed to investigate some thermodynamic properties of the static neutral AdS black hole in KR gravity. In this subsection, we will primarily focus on Case A. For detailed discussions related to Case B, please refer to the Ref. \cite{Yang:2023wtu}.

The Bekenstein-Hawking entropy is characterized by one quarter of its horizon area:
\begin{equation}
S = \frac{\mathbb{A}}{4} = \pi r_h^2,
\end{equation}
where $r_h$ is the location of the event horizon. The Hawking temperature is proportional to the surface gravity $\kappa$ on the event horizon
\begin{equation}
T = \frac{\kappa}{2\pi} = \frac{1 -\Lambda_er_h^2 +\Lambda_e\ell r_h^2}{4\pi r_h\sqrt{1-\ell}}.
\end{equation}

To compute the mass $\mathcal{M}$, the Abbott-Deser (AD) method \cite{NPB195-76} or the conformal completion method \cite{PRD73-104036} is commonly utilized.
Here, we adopt the former approach. According to the AD procedure \cite{NPB195-76}, the asymptotically AdS spacetime metric $g_{\mu\nu}$ is separated into a perturbative form:
\begin{equation}
g_{ab} = \bar{g}_{ab} +h_{ab},
\end{equation}
where the background metric $\bar{g}_{ab}$ represents the pure AdS$_4$ spacetime, and $h_{ab}$ is the perturbation part. The pure AdS metric is easily obtained by setting $M = 0$ in equation (\ref{KRAdS}) as follows:
\begin{align}
d\bar{s}_{\rm AdS}^2 =&-\left[1 -\frac{1}{3}(1-\ell)\Lambda_er^2 \right]dt^2 +\frac{1-\ell}{1 -\frac{1}{3}(1-\ell)\Lambda_er^2}dr^2 \non
&+r^2(d\theta^2 +r^2\sin^2\theta d\phi^2),
\end{align}
with the determinant being $\sqrt{-\bar{g}} = \sqrt{1-\ell}r^2\sin\theta$.

After background subtraction, each component of the perturbation metric tensor  $h_{ab}$ can be obtained, and its main asymptotic behavior at infinity is not difficult to obtain through the series expansion
\begin{equation}
\begin{aligned}
&h_{tt} \simeq \frac{2M}{r} +\mathcal{O}(r^{-2}), \qquad h_{\theta\theta} = 0, \qquad h_{\varphi\varphi} = 0,\\
&h_{rr} \simeq-\frac{(1-\ell)M[(1-\ell)\Lambda_er^2 -3]^{-1}}{(1-\ell)\Lambda_er^3 -3r +6M} +\mathcal{O}(r^{-6}) ,
\end{aligned}
\end{equation}
from which one can conclude that they are well-behaved at infinity.

Defining two symmetric tensors as did in \cite{NPB195-76}
\begin{align}
&\mathcal{K}^{abcd} =\bar{g}^{ac}\mathcal{H}^{bd} +\bar{g}^{bd}\mathcal{H}^{ac}
 -\bar{g}^{ad}\mathcal{H}^{bc} -\bar{g}^{bc}\mathcal{H}^{ad}, \\
&\mathcal{H}^{ab} = h^{ab} -\frac{1}{2}\bar{g}^{ab}h^c{}_{c}, \nonumber
\end{align}
and the conserved charges $\mathcal{Q}[\chi]$ associated with the Killing vector $\chi$ can be calculated by
\begin{equation}
\begin{aligned}
\label{ADm}
\mathcal{Q}[\chi]=&\frac{1}{16\pi}\int_0^{2\pi}d\varphi \int_0^{\pi}d\theta \sqrt{-\bar{g}}\big(\chi_c{\nabla}_d \mathcal{K}^{trcd}\\
&+\mathcal{K}^{tcdr}{\nabla}_d\chi_c\big)|_{r\to\infty},
\end{aligned}
\end{equation}
one can straightforwardly compute the conserved mass as:
\begin{equation}
\mathcal{M}=\mathcal{Q}[\pp_t] = \frac{M}{\sqrt{1-\ell}}.
\end{equation}
It is not difficult to demonstrate that the above thermodynamical quantities completely satisfy both the differential first law of black hole thermodynamics and the Bekenstein-Smarr formula
\begin{align}
d\mathcal{M} =& TdS +VdP , \\
\mathcal{M} =& 2TS -2VP ,
\end{align}
with the thermodynamic volume
\begin{equation}
V = \frac{4}{3}\pi r_h^3\sqrt{1-\ell},
\end{equation}
which is conjugate to the pressure $P = -\Lambda_e/(8\pi)$. When the effective cosmological constant $\Lambda_e$ vanishes, all the above thermodynamic formulas can consistently reduce to the static neutral black hole case in KR gravity.

\subsection{RII}
Almost fourteen years ago, it is conjectured \cite{PRD84-024037} that the AdS black hole satisfies the following reverse isoperimetric inequality (RII):
\begin{equation}
\label{rii}
\mathcal{R} = \left[\frac{(D-1)V}{\mathcal{A}_{D-2}}\right]^{1/(D-1)}
 \left(\frac{\mathcal{A}_{D-2}}{\mathbb{A}}\right)^{1/(D-2)} \ge 1 .
\end{equation}
Here, $V$ is the thermodynamic volume, $\mathbb{A} = 4S$ is the horizon area, and $\mathcal{A}_{D-2} = 2\pi^{[(D-1)/2]}/\Gamma[(D-1)/2]$. Equality is achieved for the Schwarzschild-AdS black hole, signifying that it possesses the maximum entropy. Alternatively, for a given entropy, the Schwarzschild-AdS black hole exhibits the smallest volume, endowing it with the highest efficiency in information storage.

\subsubsection{Case A}

It is straightforward to check whether the static neutral AdS black hole of Case A in KR gravity satisfies this RII or not.
It is readily known that the area of the unit two-dimensional sphere is $\mathcal{A}_2 = 4\pi$.
Substituting the thermodynamic volume: $V = 4S(\sqrt{1 -\ell}r_h)/3$ and the horizon area: $\mathbb{A} = 4S = 4\pi r_h^2$ into the above isoperimetric ratio (\ref{rii}), we now get
\begin{equation}
\mathcal{R} = \left(1 -\ell \right)^{\frac{1}{6}}.
\end{equation}
Obviously, the value range of $\mathcal{R}$ is uncertain.
If $\ell > 0$ (namely, $\xi_2 > 0$), then $\mathcal{R} < 1$. In this case, the static neutral AdS black hole in KR gravity violates the RII and is superentropic.
Otherwise, if $\ell \le 0$, one then has $\mathcal{R} \ge 1$.
In this situation, the static neutral AdS black hole in KR gravity obeys the RII and is subentropic.
Therefore, the static neutral AdS black hole in KR gravity can obey or violate the RII, and it is either subentropic or superentropic, depending upon the range of the coupling constant $\xi_2$.
This character is very similar to that of the ultraspinning (dyonic) Kerr-Sen-AdS$_4$ black holes \cite{PRD102-044007,PRD103-044014} and the six-dimensional ultraspinning Chow's black holes in gauged supergravity \cite{JHEP1121031}.

\subsubsection{Case B}
For the static neutral AdS black hole of Case B in KR gravity, the thermodynamical volume and horizon area are \cite{Yang:2023wtu}
\begin{equation}
\begin{aligned}
V = \frac{4}{3}r_hS = \frac{4\pi}{3}r_h^3 \, , \qquad \mathbb{A} = 4S = 4\pi r_h^2 \, .
\end{aligned}
\end{equation}
It is easy to obtain the isoperimetric ratio (\ref{rii}) as
\begin{equation}
\begin{aligned}
\mathcal{R} = 1 \, .
\end{aligned}
\end{equation}
Hence we have shown that the static neutral AdS black hole of Case B in KR gravity always satisfy the RII. As far as this point is concerned, the above two types of static neutral AdS black holes in KR gravity exhibit different properties. This constitutes one of the main results presented in this paper.

\section{Dynamical evolution}\label{Sec.4}
In this section, we focus on the dynamical evolution of the scalar fields in the two Cases and compare their corresponding physical properties, respectively.
The scalar fields in curved spacetime is governed by the Klein-Gordon equation, as
\begin{align}
\nabla_a\nabla^a\Psi=\frac{1}{\sqrt{-g}}\pp_a\left(g^{ab}\sqrt{-g}\pp_b\Psi\right)=0,
\end{align}
we can express the scalar fields $ \Psi $ as:
\begin{align}
\Psi(t,r,\theta,\varphi)=\psi(t,r)Y^{Lm}(\theta,\varphi),
\end{align}
and the $ Y^{Lm}(\theta,\varphi) $ denotes the scalar spherical harmonics.
By decoupling the angular section, the purely radial Klein-Gordon equation is
\begin{align}
\left(A\pp^2_r-B\pp^2_t\right)\psi-\left(\frac{rA'+4A}{2r}-\frac{AB'}{2B}\right)\pp_r\psi-\frac{AB(L^2+L)}{r^2}\psi=0,
\end{align}
where $ L $ is the azimuthal quantum number, acting as the separation constant.
It turns out that, by coordinate and function transformations, the equation of motion for the scalar field $ \Psi $ can be uniformly written as a Schr\"{o}dinger-like wave equation.
To achieve this, we set $ \psi(t,r) = r^{-1}Z(t,r) $, then
\begin{align}\label{mastereq}
\left(\partial^2_{r_*} - \partial^2_t\right)Z = \mathcal{V}Z,
\end{align}
where $ r_* $ is the standard tortoise coordinate, satisfying $ \partial_{r_*}r = \sqrt{A/B} $, and $ \mathcal{V} $ is the effective potential, given by
\begin{align}
\mathcal{V} = \frac{A'}{2rB} - \frac{AB'}{2rB^2} + \frac{A(L^2+L)}{r^2}.
\end{align}
For the convenience of numerical calculations, using the light-cone coordinates $ u = t-r_* $ and $ v = t + r_* $ \cite{PRD109-104060}, the wave equation \meq{mastereq} can be rewritten as
\begin{align}\label{masterequv0}
\left(4\pp_u \pp_v-\mathcal{V}\right)Z=0.
\end{align}
It can be effectively solved by employing numerical techniques such as the finite difference method \cite{Zhu:2014sya,Fu:2022cul,Tan:2022vfe}.
The initial distribution $ Z(u,v) $ is given by imposing the following initial conditions:
\begin{align}\label{masterequv}
Z(u,0)=0,~~~~Z(0,v)=\text{Exp}\left(-\frac{(v-v_c)^2}{2\sigma^2}\right),
\end{align}
where $ Z(0,v) $ denotes a Gaussian wave packet, localized around $v_c$ with a $\sigma$ width.
For both asymptotically flat metrics of Case A \meq{dscaseA} and Case B \meq{dscaseB}, we have uniform conditions that the Gaussian wave packet is located at $ v_c=10M $ in the Boyer-Lindquist coordinates, and its width is $ \sigma=1M $.
To simplify the computation without loss of generality, we set the black hole parameters to $ M=1 $ and neglect the effects of the cosmological constant.
We then obtain the dynamical evolution waveforms by numerically solving the partial differential equation \meq{masterequv}.
\begin{figure}[t]
\centering
\includegraphics[width=0.47\linewidth]{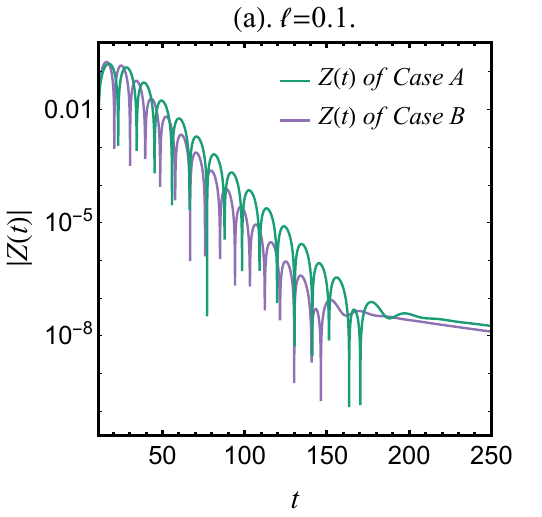}
\includegraphics[width=0.48\linewidth]{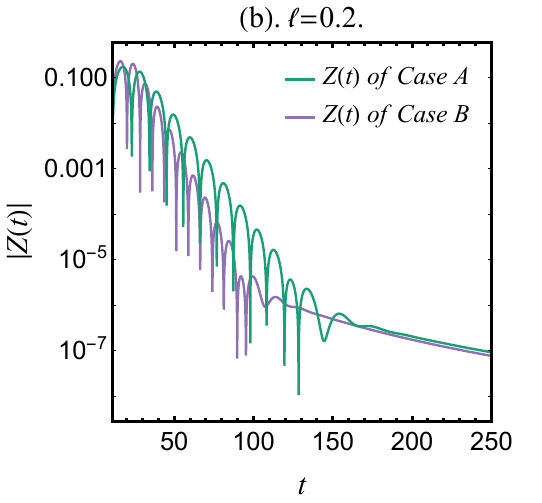}
\includegraphics[width=0.48\linewidth]{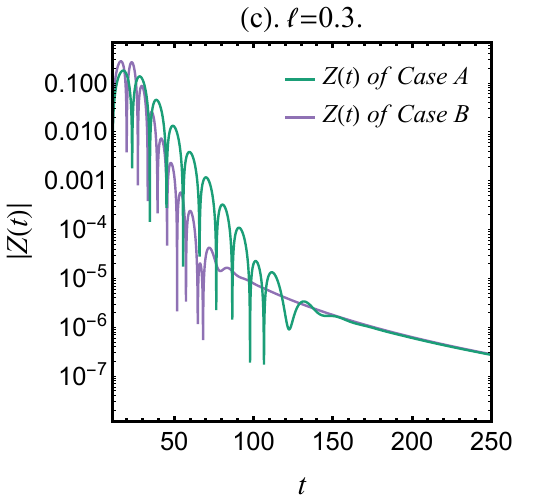}
\includegraphics[width=0.48\linewidth]{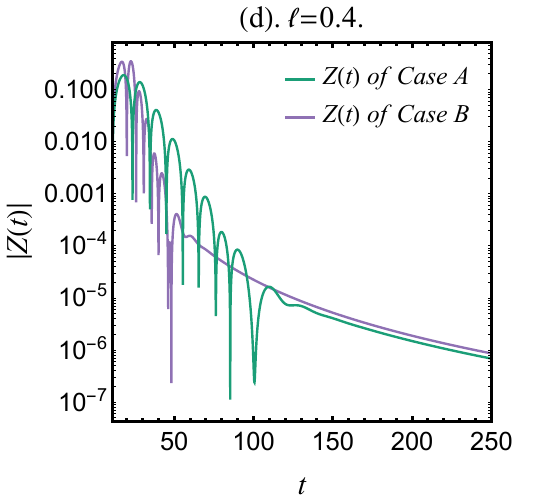}
\caption{The dynamical evolution of the wave functions $ Z(t) $ corresponds to the scalar fields in the $ L=1, n=0 $ mode.}
\label{fig2}
\end{figure}
The results of the scalar fields in both Cases are shown in Fig. \ref{fig2}, which corresponds to the azimuthal index $ L=1 $.
It is worth to note that for nonzero Lorentz violation parameters, there is a significant inconsistency between the Case A and Case B evolution waveforms.
As the Lorentz-violating parameter $ \ell $ increases, both types of scalar fields decay faster and transition from the ringdown phase to the power-law tail phase sooner.
Case B is more sensitive to the Lorentz-violating parameter than Case A.
Generally, the scalar field in Case A will have a longer lifetime.

An important aspect to consider is verifying the accuracy of the numerical method.
This verification is achieved by fitting $ Z(t,r_*) $ with a QNM frequency that includes a finite number of exponentially damped sinusoids.
For simplicity, we focus solely on the fundamental modes and omit the indices $ n $ and $ m $.
We employ a modified exponentially decaying function,
\begin{align}
Q(t)=e^{\omega_I t}A_{l}\sin(\omega_R t+B_{l}), \qquad \qquad  t\in (t_0,t_\text{end}),
\end{align}
with a fitting range from $ t_0=45M $ to $ t_\text{end}=70M $.
Here, the imaginary and real parts of the decay frequency of the waveform are represented by $ \omega_I $ and $ \omega_R $, respectively, with $ \omega=\omega_R+\omega_I $ referred to as the QNM frequency \cite{Berti2009}.
Finally, we compare the fitted results with those calculated by the continued fraction method (CFM)
\footnote{We will present the calculations of the eigenfrequencies of the scalar fields using the CFM in Appendix \ref{AppendixB}, which involves numerically solving a three-term recurrence relation.}, as presented in Table $\ref{tab1}$, where the upper and lower parts correspond to Case A and Case B, respectively.
\begin{table}[t]
\renewcommand{\arraystretch}{1.4}
\centering
\caption{Contrasting the QNM frequencys for values of $ M = 1 $ in the $ L=1, n=0 $ mode.}
\label{tab1}
\setlength\tabcolsep{2mm}{
\begin{tabular}{cccccc}
\hline\hline
\multirow{2}{*}{~\bf }	&
\multirow{1}{*}{~\bf $ \ell $~}  &
\multicolumn{1}{c}{$ M\omega  $ of CFM} & \multicolumn{1}{c}{$M\omega$ of Fitting}
\\
\hline
\multirow{3}{*}{$ \text{Case A} $}
  &$0.1 $  &0.295174-0.103087i  &0.293307-0.103813i        \\
  &$0.2 $  &0.297945-0.109519i	&0.295901-0.110013i         \\
  &$0.3 $  &0.301476-0.117314i	&0.299378-0.116572i          \\
\\
\multirow{3}{*}{$ \text{Case B} $}
  &$0.1 $  &0.345711-0.120737i	&0.346227-0.120361i
\\&$0.2 $  &0.416391-0.153058i	&0.416379-0.152833i
\\&$0.3 $  &0.514761-0.200311i	&0.522667-0.206675i
\\
\hline\hline
\end{tabular}}
\end{table}
Despite the inherent errors that can arise during numerical computations and the limitations imposed by the finite number of parameters in the fitting process, we are confident in the accuracy of the results.
This confidence is evidenced by our CFM results.

\section{Conclusions and outlooks} \label{Sec.5}

The KR gravity is an important theory that involves nonminimal coupling with an antisymmetric tensor field, resulting in spontaneous Lorentz symmetry breaking.
Our comprehensive analysis of static spherical vacuum solutions within the KR gravity model uncovers two types of black hole solutions, each exhibiting a diverging $ g_{rr} $ but finite Kretschmann scalars at a specific finite radius $ r_h $.
The first type of black hole solutions, called Case A, possesses a more general spherically symmetric form $ (-g_{tt}\ne g_{rr}^{-1}) $; in this case, the event horizon of the asymptotically flat black hole coincides with that of the Schwarzschild black hole, while for the asymptotically (A)dS black hole, it is affected by Lorentz violation.
The second type of black hole solutions, called Case B, has a specific spherically symmetric form $ (-g_{tt}=g_{rr}^{-1}) $; here, the event horizons of both asymptotically flat and (A)dS black holes are influenced by Lorentz violation.

In terms of these static neutral black holes, we have investigated the following properties:
\begin{itemize}[leftmargin=*,noitemsep,topsep=0pt,partopsep=0pt]
\item \textbf{Thermodynamical quantities:}
The thermodynamic quantities of the static neutral AdS black hole in Case A in the KR gravity were calculated and obeyed both the first law and the Bekenstein-Smarr mass formulas.
\item \textbf{RII:}
The static neutral AdS black hole of Case A in the KR gravity, which can violate the RII, marks the first discovery of a black hole in static AdS spacetime that can be superentropic due to Lorentz violation; however, Case B always obeys the RII.
\item \textbf{Dynamical evolution:}
The dynamical evolution waveform of scalar fields in static neutral KR spacetimes decays faster and has a shorter lifespan compared to Schwarzschild spacetime, and Case B is more significantly affected.
\end{itemize}

There are three promising further topics to be pursued in the future.
Although the black hole metrics of Case A in the KR gravity theory have a form similar to those of the static black hole in bumblebee gravity theory \cite{Casana2018,Maluf2021}, their gravitational perturbation master equations will be entirely different because they originate from different field equations.
Based on our previous work on black hole perturbations in bumblebee gravity \cite{Liu2023}, we conclude that the gravitational perturbation isospectrality is broken \cite{Liu:2022dcn}. However, for the KR gravity theory, this remains an open question.
Furthermore, constructing rotating black hole solutions is an intriguing aspect of the KR gravity theory.
When black holes have angular momentum, they exhibit additional properties such as superradiant instability \cite{Berti2009} and deformation of the black hole shadow \cite{Liu:2024lbi,Chen:2023wna,Zhong:2021mty,Guo:2018kis}.
Nevertheless, solving the gravitational field equations to obtain an exact rotating solution is very challenging.
One can consider the slow rotation approximation \cite{Pani2012prl,Pani2013prl}, which greatly reduces the difficulty of the solution while still accurately reflecting the properties of black holes with angular momentum. For more details, please see our separate work \cite{LWW2024} about the slow rotation extension of the present paper.
In addition, exploring the horizon geometry \cite{PRD103-104020,PRD104-L121501}, topological classifications \cite{Wei:2022dzw,Wu:2022whe,Wu:2023sue,Wu:2023xpq,Wu:2023fcw,
Wu:2023meo,Zhu:2024jhw}, and phase transition criticality \cite{Hawking:1982dh,
Kubiznak:2012wp,Ahmed:2023dnh,Wu:2024rmv} of the static neutral black holes in KR gravity found in this paper is expected to significantly advance our understanding of the intrinsic nature of gravity.

\acknowledgments
We are greatly indebted to the anonymous referee for his/her constructive comments to improve the presentation of this work. We also thank Professor Shuang-Qing Wu for useful discussions. This work is supported by the National Natural Science Foundation of China (NSFC) under Grants No. 12205243, No. 12375053, No. 12122504, and No. 12035005; the Sichuan Science and Technology Program under Grant No. 2023NSFSC1347; the Doctoral Research Initiation Project of China West Normal University under Grant No. 21E028;
the innovative research group of Hunan Province under Grant No. 2024JJ1006; and the Hunan Provincial Major Sci-Tech Program under Grant No.2023ZJ1010.

\appendix

\section{The covariant derivative of KR fields}\label{AppendixA}
In this appendix, we express the covariant derivative terms of the KR field in Eq. \meq{EQB} using common derivatives:
\begin{equation}
\begin{aligned}
&\nabla_c\nabla_{a}\beta^{cd}\beta_{bd}=\Gamma^{c}{}_{ad} \Gamma^{e}{}_{ce} \beta_{bf} \beta^{df} -  \Gamma^{c}{}_{ad} \Gamma^{d}{}_{ce} \beta_{bf} \beta^{ef} \\
&+ \Gamma^{c}{}_{ad} \Gamma^{d}{}_{be} \beta_{cf} \beta^{ef}  -  \Gamma^{c}{}_{ab} \Gamma^{d}{}_{de} \beta_{cf} \beta^{ef} + \Gamma^{c}{}_{cd} \beta^{de} \partial_{a}\beta_{be}\\
& -  \Gamma^{c}{}_{bd} \beta^{de} \partial_{a}\beta_{ce} + \Gamma^{c}{}_{cd} \beta_{be} \partial_{a}\beta^{de} -  \Gamma^{c}{}_{bd} \beta_{ce} \partial_{a}\beta^{de} \\
&-  \beta_{ce} \beta^{cd} \partial_{d}\Gamma^{e}{}_{ab} -  \partial_{a}\beta^{cd} \partial_{d}\beta_{bc} -  \partial_{a}\beta_{bc} \partial_{d}\beta^{cd} \\
& -  \beta^{cd} \partial_{d}\partial_{a}\beta_{bc} -  \beta_{bc} \partial_{d}\partial_{a}\beta^{cd} -  \beta_{bc} \beta^{cd} \partial_{e}\Gamma^{e}{}_{ad} \\
&+ \Gamma^{c}{}_{ab} \beta^{de} \partial_{e}\beta_{cd} + \Gamma^{c}{}_{ab} \beta_{cd} \partial_{e}\beta^{de},
\end{aligned}
\end{equation}
\begin{equation}
\begin{aligned}
&\nabla^c\nabla_c\beta_a^{~d}\beta_{bd}=\Gamma^{c}{}_{be} \Gamma^{ed}{}_{d} \beta_{a}{}^{f} \beta_{cf} + \Gamma^{c}{}_{ae} \Gamma^{ed}{}_{d} \beta_{bf} \beta_{c}{}^{f}\\
& + \Gamma^{c}{}_{b}{}^{e} \Gamma^{d}{}_{ce} \beta_{a}{}^{f} \beta_{df} + \Gamma^{c}{}_{a}{}^{e} \Gamma^{d}{}_{be} \beta_{c}{}^{f} \beta_{df} + \Gamma^{c}{}_{ae} \Gamma^{d}{}_{b}{}^{e} \beta_{c}{}^{f} \beta_{df} \\
&+ \Gamma^{c}{}_{a}{}^{e} \Gamma^{d}{}_{ce} \beta_{bf} \beta_{d}{}^{f} -  \Gamma^{ce}{}_{e} \beta_{bd} \partial_{c}\beta_{a}{}^{d} -  \Gamma^{ce}{}_{e} \beta_{a}{}^{d} \partial_{c}\beta_{bd} \\
& + \beta_{bc} \beta^{c}{}_{e} \partial^{d}\Gamma^{e}{}_{ad} + \beta_{a}{}^{c} \beta_{ce} \partial^{d}\Gamma^{e}{}_{bd} -  \Gamma^{c}{}_{b}{}^{e} \beta_{cd} \partial_{e}\beta_{a}{}^{d} \\
&-  \Gamma^{c}{}_{a}{}^{e} \beta_{c}{}^{d} \partial_{e}\beta_{bd} -  \Gamma^{c}{}_{b}{}^{e} \beta_{a}{}^{d} \partial_{e}\beta_{cd} -  \Gamma^{c}{}_{a}{}^{e} \beta_{bd} \partial_{e}\beta_{c}{}^{d}\\
& + \partial_{e}\beta_{bc} \partial^{e}\beta_{a}{}^{c} -  \Gamma^{c}{}_{be} \beta_{cd} \partial^{e}\beta_{a}{}^{d} + \partial_{e}\beta_{a}{}^{c} \partial^{e}\beta_{bc} \\
&-  \Gamma^{c}{}_{ae} \beta_{c}{}^{d} \partial^{e}\beta_{bd} -  \Gamma^{c}{}_{be} \beta_{a}{}^{d} \partial^{e}\beta_{cd} -  \Gamma^{c}{}_{ae} \beta_{bd} \partial^{e}\beta_{c}{}^{d}\\
& + \beta_{bc} \partial^{e}\partial_{e}\beta_{a}{}^{c} + \beta_{a}{}^{c} \partial^{e}\partial_{e}\beta_{bc}.
\end{aligned}
\end{equation}
Here, the christoffel symbols are
\begin{equation}
\begin{aligned}
\Gamma^{a}{}_{bc}&=\frac{1}{2}g^{ad}\left(\pp_b g_{cd}+\pp_c g_{bd}-\pp_d g_{bc}\right),\\
\Gamma^{ab}{}_{c}&=\frac{1}{2}g^{ad}g^{be}\left(\pp_cg_{ed}+\pp_eg_{cd}-\pp_dg_{ec}\right),\\
\Gamma^{a}{}_{b}{}^{c}&=\frac{1}{2}g^{ad}g^{ce}\left(\pp_bg_{ed}+\pp_eg_{bd}-\pp_d g_{be} \right).
\end{aligned}
\end{equation}

\section{CFM for the scalar fields}\label{AppendixB}
To verify the accuracy of the numerical method in Sec. \ref{Sec.4}, we will use the CFM to calculate, as reference data, the eigenfrequencies (i.e., QNMs) of the scalar field equation \meq{mastereq}, following the Leaver method \cite{Leaver1985,Leaver:1986gd,Pani2013IJMPA,Liu:2023MOG}.
We need to transform the problem from the time domain to the frequency domain by setting $ Z(t,r)=e^{-i\omega t}Z(r) $.
For both Case A ($s = 0$) and Case B ($s = 1$), Eq. \meq{mastereq} can be uniformly rewritten as
\begin{align}\label{ZX}
\pp_x^2Z_s+\left[(1-\ell)^{s+1}\omega^2-\mathcal{V}_s\right]Z_s=0,
\end{align}
where $ x=\int F_s^{-1}dr $ with $ F_s=1-2M(1-\ell)^s/r $, and the specific form of the effective potential is:
\begin{align}
\mathcal{V}_s=F_s\left(\frac{(1-\ell)L(L+1)}{r^2}+\frac{2M(1-\ell)^s}{r^3}\right).
\end{align}
According to the boundary conditions at the horizon and at infinity based on Eq. \meq{ZX}, the wave function $ Z_s $ exhibits the following asymptotic behaviors:
\begin{equation}\label{solveeq}
Z_s \sim
\begin{cases}
e^{-i\sqrt{(1-\ell)^{s+1}}\omega x} & \text{for}~~r \rightarrow r_h,\\
e^{i\sqrt{(1-\ell)^{s+1}}\omega x} & \text{for}~~r \rightarrow \infty,
\end{cases}
\end{equation}
which indicates a pure-ingoing wave at the event horizon and an outgoing wave at infinity.
The asymptotic form of the fields as $ r_*\rightarrow\infty $ can be rewritten as
\begin{align}\label{psiinf}
e^{i\sqrt{(1-\ell)^{s+1}}\omega x}=e^{i\sqrt{(1-\ell)^{s+1}}\omega r} (r/r_h)^{i\sqrt{(1-\ell)^{s+1}}\omega r_h},
\end{align}
The wave functions $ Z_s $ that expands at the black hole horizon can be written in the following form:
\begin{equation}\label{Leaverpsi}
Z_s=e^{i\sqrt{(1-\ell)^{s+1}}\omega x}
\left(1-\frac{r_h}{r}\right)^{-i\sqrt{(1-\ell)^{s+1}}\omega r_h}
\sum_{n=0}^{\infty}a_n\left(1-\frac{r_h}{r}\right)^n.
\end{equation}
Combining Eqs. \meq{ZX} and \meq{Leaverpsi}, the expansion coefficients are defined by a three-term recursion relation
\begin{equation}
\begin{aligned}\label{ditui}
\alpha_0a_1&+\beta_0a_0=0,\\
\alpha_na_{n+1}+\beta_na_n&+\gamma_na_{n-1}=0,~~~ n>0.
\end{aligned}
\end{equation}
The recurrence coefficients $ \alpha_n $, $ \beta_n $, and $ \gamma_n $ are simple functions of $ n $ and black hole parameters.
Their explicit forms are given by:
\begin{equation}
\begin{aligned}
\alpha_n &= n \left[ n - 2i r_h \sqrt{(1-\ell)^{s+1}} \omega \right], \\
\beta_n &= 8r_h^2 (1-\ell)^{s+1} \omega^2 - 2n^2 - L(L+1)(1-\ell) \\
& \quad + (1+2n) \left[ 4i r_h \sqrt{(1-\ell)^{s+1}} \omega - 1 \right], \\
\gamma_n &= \left[ 1 + n - 2i r_h \sqrt{(1-\ell)^{s+1}} \omega \right]^2.
\end{aligned}
\end{equation}
Note that when $ \ell=0 $, the recurrence relation \meq{ditui} returns to the Schwarzschild case.
The power series in Eq. \meq{Leaverpsi} converges for $ r > r_h $, implying that the eigenfunction satisfies the QNM boundary condition as defined by Eq. \meq{solveeq}.
This convergence condition implies that we only need to provide the initial value of $ a_0 $ and recursively calculate for sufficiently large values of $ n $, where the following condition holds,
\begin{align}\label{ana1}
a_{n}(\omega)-a_{n-1}(\omega)=0.
\end{align}
The eigenfrequencies $ \omega $ can be readily determined by solving Eq. \meq{ana1}, with the initial value $ a_0=1 $ set without loss of generality.
~~\\~~\\~~\\~~

\end{document}